\documentclass[10pt,twocolumn,journal]{IEEEtran}

\usepackage[top=1in, bottom=0.75in, left=0.75in, right=0.75in]{geometry}
\usepackage{graphicx}
\usepackage{amssymb}
\usepackage{epsfig}
\usepackage{subcaption} 
\usepackage{epstopdf}
\usepackage{algorithm}
\usepackage{algorithmic}
\usepackage{amsmath}
\usepackage{amsfonts}
\usepackage{dsfont}
\usepackage{url}
\usepackage{chemarr}
\usepackage{chemarrow}
\usepackage{bbold}
\usepackage[version=3]{mhchem}
\usepackage{mathabx}
\usepackage{xcolor}
\usepackage[numbers,sort&compress]{natbib}
\usepackage{comment}
\usepackage{authblk}

\newtheorem{remark}{Remark}
 
\graphicspath{{./img}{./}}

\begin{document}
\title{A New Architecture for Energy Efficient Fault Detection Using Energy Harvesters}
\author[1]{Dongti Zhang}
\author[2]{Patricio Peralta-Braz}
\author[1]{Chun Tung Chou}
\author[2]{Elena Atroshchenko}
\author[2]{Mehrisadat Makki Alamdari}
\author[1]{Mahbub Hassan\thanks{* mahbub.hassan@unsw.edu.au}}
\affil[1]{School of Computer Science and Engineering, University of New South Wales, Sydney, Australia}
\affil[2]{School of Civil and Environmental Engineering, University of New South Wales, Sydney, Australia}

\renewcommand\Authands{ and }

\maketitle

\begin{abstract}
The current battery-powered fault detection system for vibration monitoring has a rather limited lifetime. This is because the high-frequency sampling (typically tens of kilo-Hertz) required for vibration monitoring results in high energy consumption in both the analog-to-digital (ADC) converter and wireless transmissions. This paper proposes a new fault detection architecture that can significantly reduce the energy consumption of the ADC and wireless transmission. Our inspiration for the new architecture is based on the observation that the many tens of thousand of data samples collected for fault detection are ultimately transformed into a small number of features. If we can generate these features directly without high frequency sampling, then we can avoid the the energy cost for ADC and wireless transmissions. We propose to use piezoelectric energy harvesters (which can be designed to have different frequency responses) and integrators to obtain these features in an energy-efficient manner.  
By using a publicly available data set for ball bearing fault detection (which was originally sampled at 51.2kHz) and piezoelectric energy harvester models, we can produce features, which when sampled at 0.33Hz, give a fault detection accuracy of 89\% while reducing the sampling requirement by 4 orders-of-magnitude. 
\end{abstract}
   
\section{Introduction}
\label{sec:intro}

Mechanical systems are employed in a wide range of industries. Machinery is vital of driving industrial processes. Its reliability and efficiency are the fundamental to enhance productivity and improving quality of life \cite{Randall2010,Wen2018}. Due to harsh working environments and complex operating conditions \cite{Yao2023}, key components of mechanical equipment are prone to damage, leading to performance degradation and even equipment failure. Health monitoring and fault detection are essential to ensure the safe operation of manufacturing system \cite{Cerrada2018}. Recently, the emergence of advanced technologies, including wireless sensor networks, machine learning and neural networks has accelerated the achievement of high performance machine fault detection \cite{Alsheikh2014,Lei2020}.

The current architecture for data collection for fault diagnosis follows the standard architecture for digital data acquisition where sensor signals are digitised by analog-to-digital (ADC) converters and the sampled data are then transferred wirelessly to a server for further processing. For precise analysis of the machine’s condition, the ADCs are expected to collect vibration samples with a high sampling rate in the order of tens of kilo-Hertz~\cite{havinga}. This high sampling rate not only consumes high ADC energy~\cite{ADC}, it also elevates the energy usage of wireless transmission due to the larger volume of data generated for transmission. This in turns leads to frequent battery replacements which adds to operating costs as batteries have to be changed manually \cite{xidas:iot:lifetime}. Moreover, the production and disposal of batteries have an impact on the environment. In this paper, we propose a new architecture for data collection to address the bottleneck of requiring high frequency sampling for monitoring machine vibrations. 

\begin{figure}[t]
    \centering
    \begin{subfigure}[t]{0.5\textwidth}
        \centering
        \includegraphics[trim=0.5cm 11.5cm 6cm 5.2cm, clip=true, width=\columnwidth]{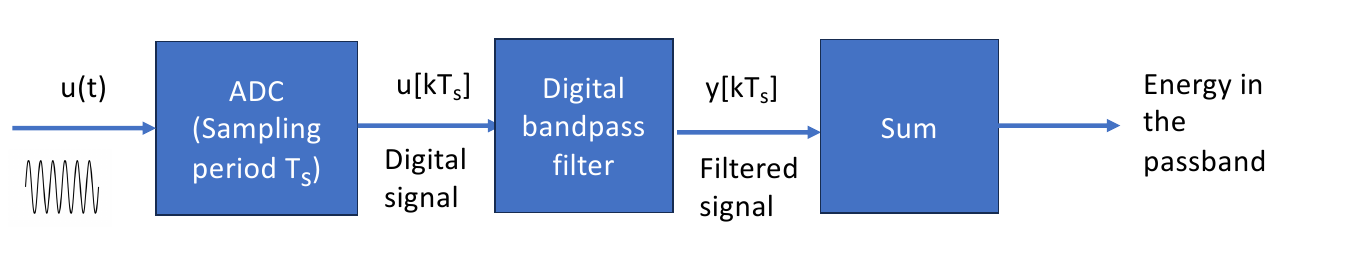}
        \caption{}
      \label{fig:bp_filter:1}
    \end{subfigure} 
    
    \begin{subfigure}[t]{0.5\textwidth}
        \centering
        \includegraphics[trim=0.5cm 11.5cm 6cm 5.2cm, clip=true, width=\columnwidth]{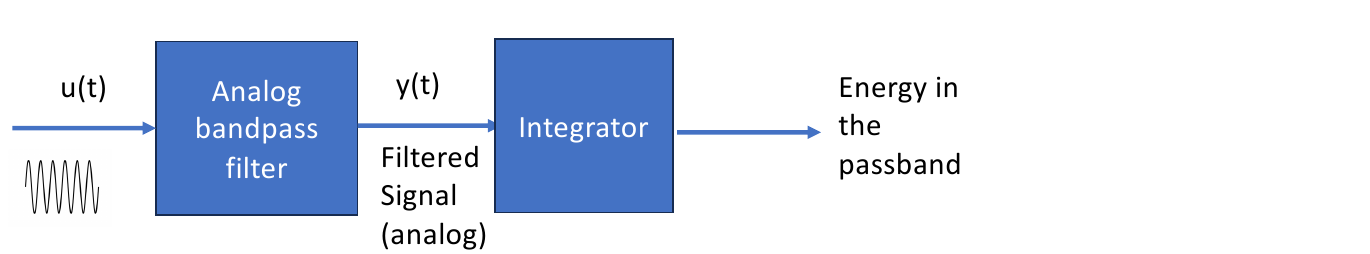}
        \caption{}
        \label{fig:bp_filter:2}
      \end{subfigure}  
      \caption{Digital versus analog computation.}
      \label{fig:bp_filter}
\end{figure}

An important requirement for the success of fault detection is the availability of informative features to differentiate between the various machine states. For many fault detection problems, especially for those on monitoring machine vibration, frequency domain features are often useful \cite{Pandarakone:ECCE:2018}. For example, if the required frequency domain feature is the energy of the vibration signal in a given frequency band, then one may proceed as in Fig.~\ref{fig:bp_filter:1} where the vibration signal $u(t)$ (which can be acceleration or other physical measurements related to vibration) is first sampled, then the digitised signal is passed through a digital band-pass filter, and then the energy in the frequency band is determined by summing up the digital filtered output. The drawback of this architecture, as pointed out in the last paragraph, is its high energy consumption. In this paper, we propose an alternative architecture whose aim is to reduce the energy consumption in obtaining features for fault detection. Our proposal is based on using analog computation. Let us continue on our earlier example, the energy of signal in a frequency band can be computed in a analog manner by first passing the the signal through an analog filter and then an integrator, as in Fig.~\ref{fig:bp_filter:2}. In this paper, we propose a new architecture for fault detection which uses piezoelectric energy harvesters (PEHs) as analog filters and integrators to compute the output energies of these PEHs. This new architecture has a number of advantage. First, the analog computation being carried out by the piezoelectric harvesters is self-powered. Second, we use the integrators to help us to reduce the sampling requirement.

We use a publicly available dataset \cite{Thuan2023} on ball bearing fault detection (which is originally sampled at 51.2kHz) to demonstrate the feasibility of our proposal. By using the dataset together with models of piezoelectric harvesters, we are able to show that we can use features collected at 0.33Hz to achieve a 89\% fault detection accuracy. 

The contributions of this paper are:
\begin{itemize}
    \item We propose a new architecture for fault detection which uses piezoelectric harvesters and integrators as a way to directly extract features for fault detection. 
    \item As a proof of concept, we use a publicly available dataset to demonstrate that our proposed architecture can achieve a 89\% accuracy of fault detection. Furthermore, the energy burden on the ADC and wireless transmissions is significantly reduced as we can now sample at 0.33Hz rather than the original frequency of 51.2kHz. 
\end{itemize}

\section{Contrasting the existing and the proposed fault detection architectures}
\label{sec:contrasting}
In this section, we first describe the current data collection architecture for fault detection. We then present a thought experiment to explain the rationale behind our proposed architecture before describing it. 

\subsection{Current architecture for fault detection}
\label{sec:existing_arch}

\begin{figure}[t]
\centering
\includegraphics[page=1,trim=1.5cm 11cm 15.5cm 2.5cm, clip=true, width=1.0\columnwidth]{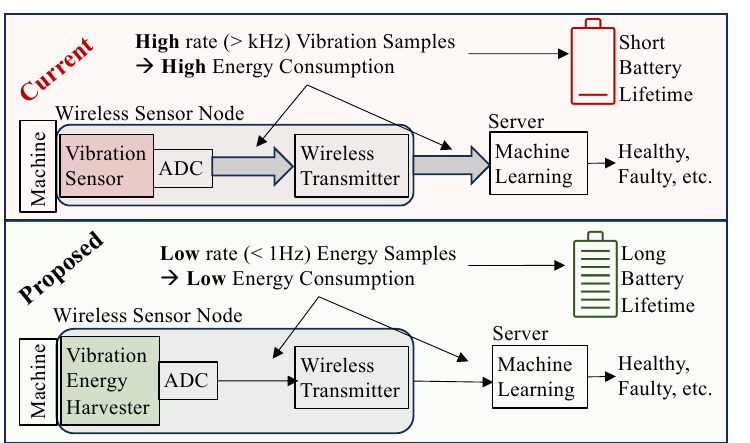}
\caption{Current architecture versus proposed architecture.}
\label{fig:current_vs_proposed}
\end{figure}

Fig.~\ref{fig:current_vs_proposed} depicts a very commonly used architecture for fault detection \cite{DAS}. The architecture consists of these components: battery, sensor, ADC, wireless transmission and machine learning. The sensor is attached to the machine to be monitored, e.g., acceleration sensors are commonly used to monitor vibrations. The sensor signal is then digitised by an ADC. The digitised data are then transmitted over a wireless interface to a server that runs the machine learning algorithm which decides whether a fault has occurred or not. 

The two most energy hungry components, which are used for data collection, in current architecture are the ADC and the wireless transmissions \cite{Lan2020}.  The energy consumption of these two components are roughly proportional to the sampling frequency \cite{mulleti2023powerefficient} as higher sampling frequency produces more data samples. However, there are obstacles to lowering the sampling frequency. Fundamentally, according to the Shannon/Whittaker sampling criterion \cite{Mallat:Wavelet}, sampling frequency must be at least two times of the highest frequency component in the signal to be sampled. However, in practice, the sampling frequency used is a few times higher than this lower limit. For the monitoring of vibration signals, a sampling frequency of tens of kHz is quite typical. 

There is much effort in the literature to reduce the energy consumption of the data collection by reducing the energy consumption in wireless transmissions \cite{Khalifeh:JIII:2023} or by duty cycling \cite{Chu:CompNet:2023}. However, there does not appear to be effort to address the energy consumption bottleneck due to high frequency sampling. In the next section, we will describe a new fault detection architecture that can significantly reduce the energy consumption. 

\subsection{Proposed fault detection architecture}
\label{sec:proposed_arch}
Our proposed architecture, which is shown in the lower half of Fig.~\ref{fig:current_vs_proposed}, is based on vibration energy harvester. If we compare the current and proposed architectures, the only component change is to replace vibration sensors by vibration energy harvesters. 

\subsubsection{Vibration energy harvesters}
We will begin our discussion by having a deeper look into vibration energy harvesters. In order to make this discussion concrete, we will use piezo-electric energy harvesters (PEHs) as the energy harvester. However, in principle, we can use other types of energy harvesters provided that they can provide frequency selective response\footnote{An input-output device is said to have frequency selective behaviour if it allows certain frequencies at the input to pass through the device to reach the output while stopping or strongly attenuating other frequencies. Typical frequency selective behaviours include low-pass, high-pass, band-pass, multiple pass bands etc.} to vibration signals. 

Although PEHs have been viewed primarily as strain-to-voltage transducers, there is a new realisation that PEHs can also provide useful information on the system that they are attached to. In our previous work in \cite{Ma2020}, we realised gait recognition by using the rectified voltage signals from two PEHs which are attached to the shoes worn by the human subjects. In another of our work \cite{Lan2020}, PEH signals are used to classify five different daily activities with a high accuracy while consuming less system power compared to conventional approaches. This shows that PEHs can also be used as a sensor. In fact, we have recently advocated that PEHs should be viewed simultaneously as an energy harvester and a sensor \cite{Peralta:TITS}. In addition, PEHs do not require battery power to produce its voltage signal, so one can view them as self-powered sensors. 

\begin{center}
    \begin{figure}[h]
        \centering
        \includegraphics[width=0.45\textwidth]{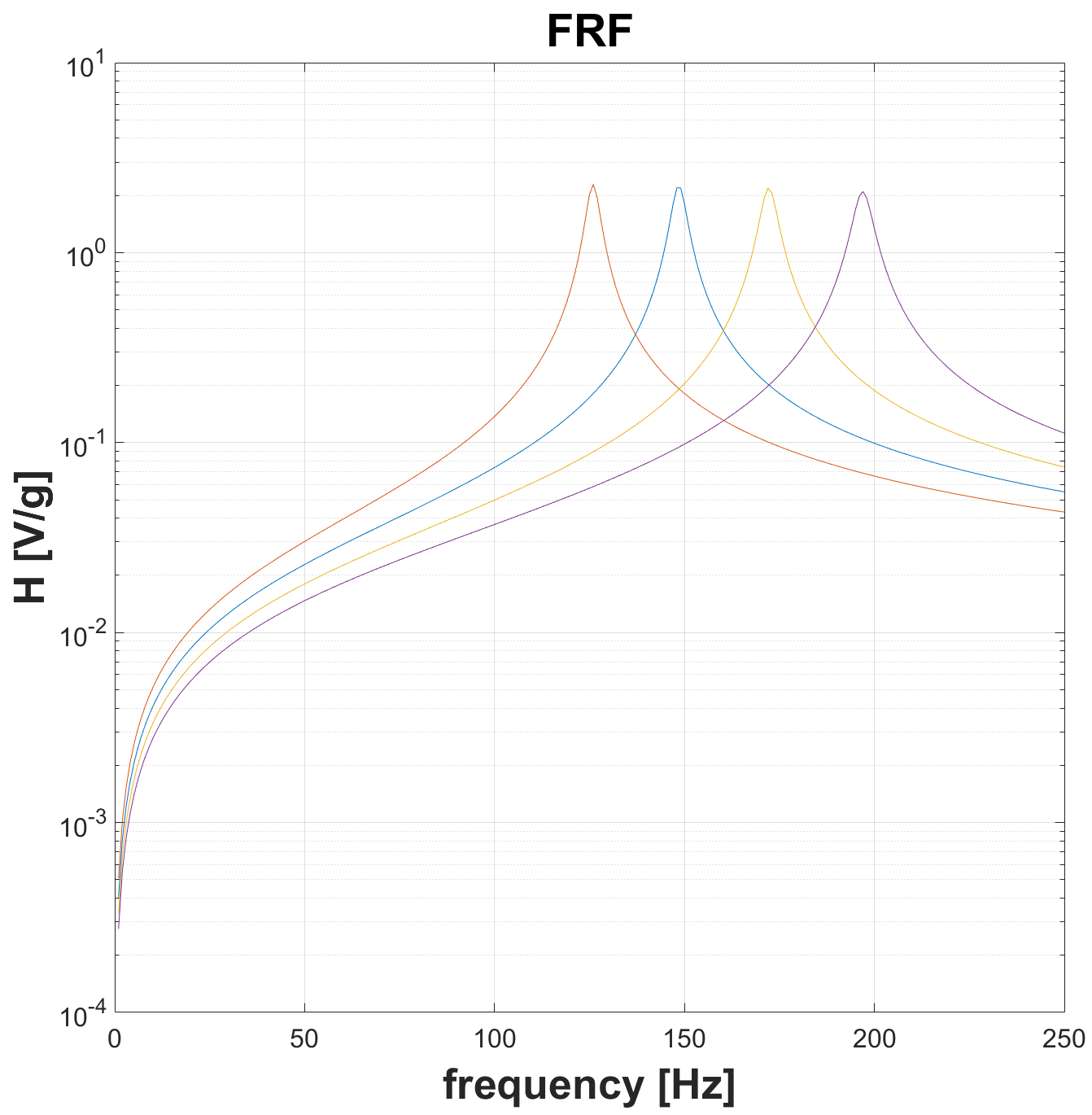}
        \caption{FRFs of four different design PEHs with PZT thickness 0.35mm, 0.4mm, 0.45mm and 0.5mm. The simulation assumes the input is acceleration (which is normalised with respect to the acceleration due to gravity $g$) and the output is voltage (in Volts).}
        \label{fig:peh:frf}
    \end{figure}
\end{center}

An interesting aspect of PEHs is that they can be made to produce frequency selective response to mechanical strain. The frequency selective response of a PEH can be characterised by the magnitude of frequency response function (FRF). Fig.~\ref{fig:peh:frf} shows the magnitude of FRFs $|H(j 2\pi f)$| against frequency $f$ of four PEHs with four different sets of form factors calculated by using the model in \cite{Peralta:JSV:2020,Peralta:MSSP:2023}. In particular, we assume that lead zirconate titanate (PZT) is the substrate for the PEHs and vary its thickness from 0.35mm to 0.5mm with a step of 0.05mm. We can see that these PEHs have peak frequency responses at about 125Hz, 150Hz, 175Hz and 200Hz. These FRFs show band-pass characteristics with a 3dB-bandwidth of about 10Hz. We will make use of this frequency-selective property of PEHs for energy efficient fault detection, which will be discussed next. 

\subsubsection{Rationale on using PEHs}
\label{sec:rationale}

We will use a thought experiment to motivate the proposed fault detection architecture. We consider the problem of monitoring the state of a machine which vibrates. We assume that when the machine is healthy, it vibrates at 200Hz. However, when the machine is faulty, its vibration frequency changes to 150Hz. We assume that we have two PEHs with band-pass behaviour. One has a pass-band of 195-205Hz while the other PEH's band-pass is 145-155Hz. The thought experiment here is admittedly simplistic because the machine state is characterised by a single frequency. We will use real data in Sec.~\ref{sec:evaluation} for a more realistic evaluation. 

\begin{figure*}[t]
    \centering
    \begin{subfigure}[t]{\textwidth}
        \centering
        \includegraphics[trim=0.5cm 4cm 0.0cm 5.2cm, clip=true, width=0.9\textwidth]{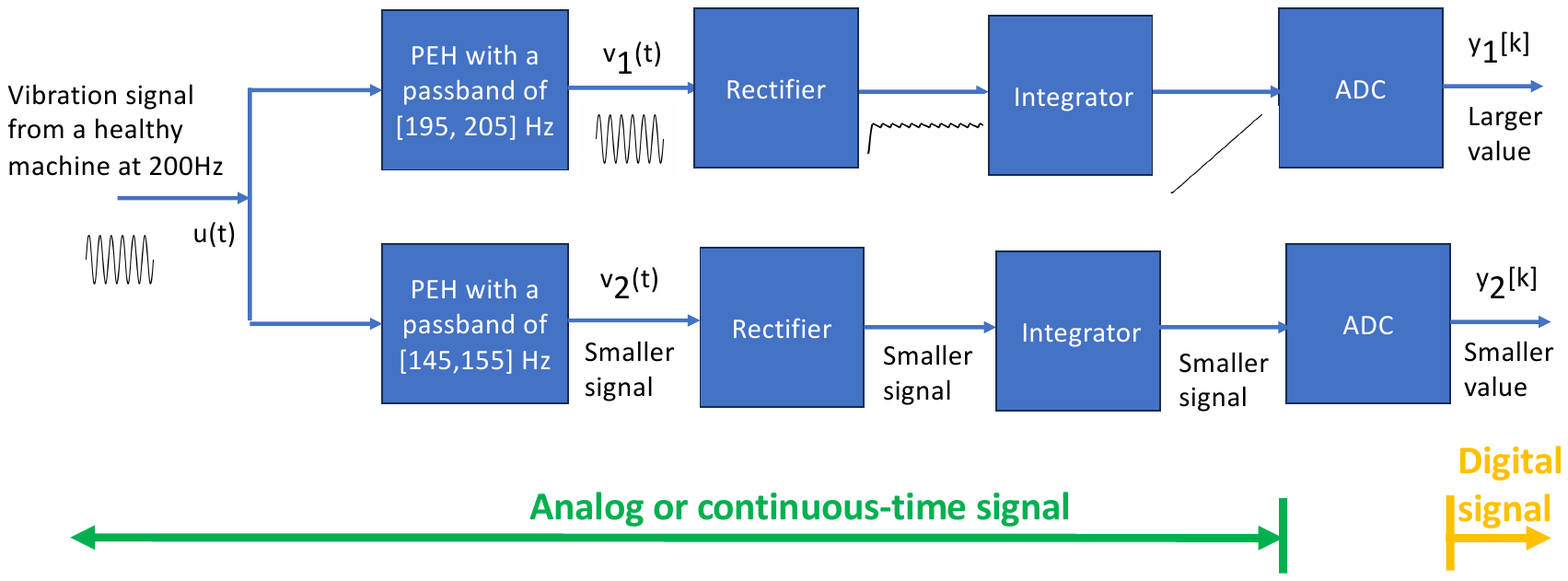}
        \caption{}
      \label{fig:thought:1}
    \end{subfigure} 
    
    \begin{subfigure}[t]{\textwidth}
        \centering
        \includegraphics[trim=0.5cm 4cm 0.0cm 5.2cm, clip=true, width=0.9\textwidth]{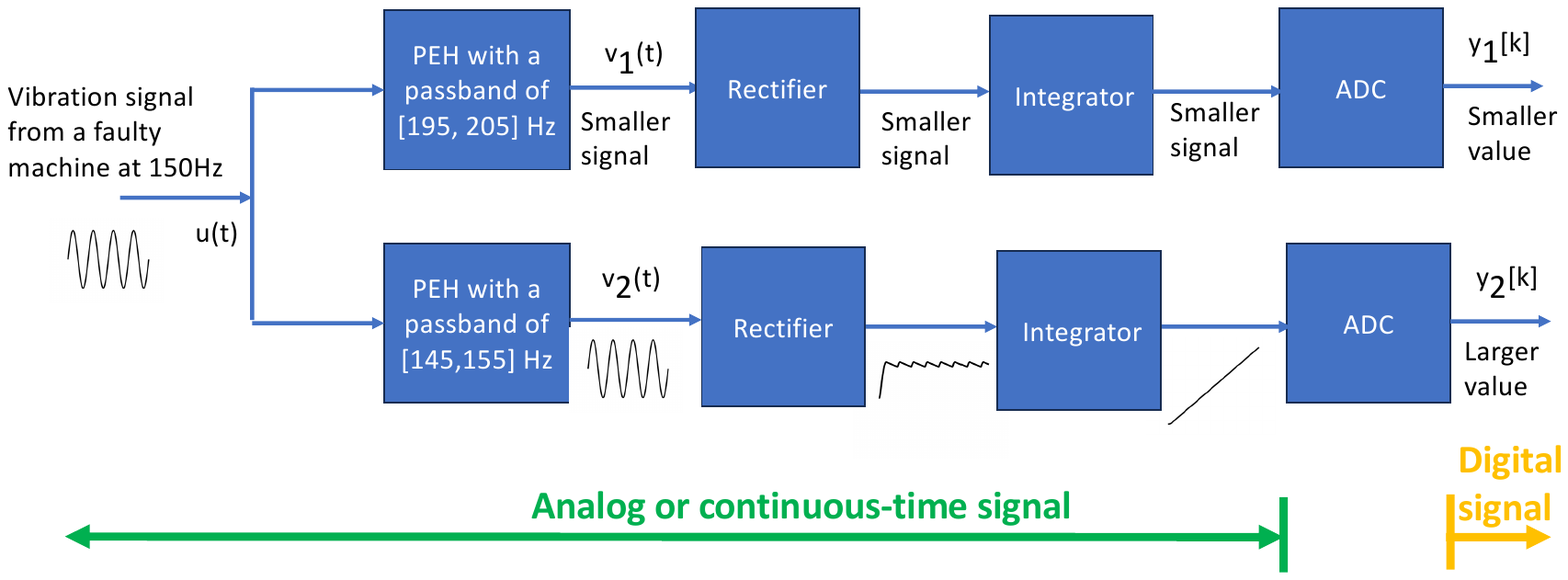}
        \caption{}
        \label{fig:thought:2}
      \end{subfigure}  
      \caption{Thought experiment on the proposed architecture.}
      \label{fig:thought}
\end{figure*}

We first consider the case where the machine is in a healthy state. The 200Hz vibration signal of the healthy machine will generate voltage signals $v_1(t)$ and $v_2(t)$ at the two PEHs as shown in Fig.~\ref{fig:thought:1}. However, the amplitudes of the voltage signals of the two PEHs are expected to be different because the two PEHs have different pass-bands. We expect $v_1(t)$ to have a higher amplitude compared to $v_2(t)$. 

We remark that the signals $v_i(t)$ ($i$ = 1, 2) are continuous-time signal. Formally, if $u(t)$ is the continuous-time vibration signal and $h_i(t)$ are the impulse response of the two PEHs, then $v_i(t) = h_i(t) \ast u(t)$ where $\ast$ is the convolution operator. The purpose of this remark is to emphasize that the signals $v_i(t)$ are obtained via analog filtering by the PEHs and no sampling has yet taken place. 

As a continuation of our thought experiment, we next consider the case when the machine is in a faulty state where the machine vibrates at 150Hz. In Fig.~\ref{fig:thought:2}, we consider what happens when this 150Hz signal passes through the two frequency-selective PEHs in Fig.~\ref{fig:thought:2}. We can see that, when the machine is faulty, we expect $v_1(t)$ to have a lower amplitude compared to $v_2(t)$. 

The above discussion points to a possible method to differentiate whether the machine is in the healthy or faulty state by comparing the amount of energy in the signals $v_1(t)$ and $v_2(t)$. If the energy in $v_1(t)$ is a lot higher (resp. lower) than that in $v_2(t)$, then we can decide that the machine is in the healthy (faulty) state. We remark that both rectifier and integrator are analog circuits, so the outputs of these two processing blocks are continuous-time signals, as indicated in Fig.~\ref{fig:thought}.

Since we know that the information is contained in the energies in the voltages $v_1(t)$ and $v_2(t)$, we therefore use an rectifier and integrator (which can be realised by a capacitor) to capture the energy in the signal. Our proposal is to measure the amount of energy captured periodically by using a sampler with a sampling period of $T$, which is a design parameter. 

Fig.~\ref{fig:thought} show that we pass $v_i(t)$ ($i = 1,2$) through a rectifier, integrator and sampler to obtain the sampled signals $y_i$. Let the sampling time instants be $kT$ where $k = 0, 1, 2, ...$. We can write the output samples $y_i[k]$ as:
\begin{equation}
y_i[k] = \int_{(k-1)T}^{kT} \frac{v_i(t)^2}{R} dt.     
\label{eq:int_and_sample}
\end{equation}
This means $y_i[k]$ is the energy produced by the PEH in the time interval $[(k-1)T,kT]$ based on a resistance of $R$ Ohms. 

We will choose a sampling time $T \gg \frac{1}{f_m}$ where $f_m$ is the maximum frequency in the pass-band so that the integration is carried out over many cycles of $v_i(t)$. This is so that $y_i[k]$ is a low frequency signal. 

Based on the discussion in the last paragraphs, we can see that we can use the relative values $y_1[k]$ and $y_2[k]$ to determine whether the machine is in healthy or faulty state. In addition, the choice of $T \gg \frac{1}{f_m}$ means the sampling requirement of the proposed architecture is a lot lower than that of the current architecture. This means we can significantly reduce the burden on ADC and wireless transmissions. This in turns means that we can achieve fault detection with a much smaller energy footprint. 

\subsubsection{Discussion}
The key idea behind our proposed architecture is to use PEHs as self-powered signal processing units that can provide filtered vibration signals to characterise the machine state. This extends the current thinking in the literature which views PEHs either as an energy source or an energy-source-plus-sensor to a \textsl{novel} viewpoint that PEHs are energy source, sensor and analog signal processor. To the best of our knowledge, the idea of using PEHs for analog signal processing has not been considered before. 

Our proposal to use PEHs as an analog signal processing device has also come at a time when computing researchers have started to realise that analog computer and communications can be useful to overcome some of the limitations of digital computing and communications. For example, analog circuits can be more energy efficient \cite{Kalofonou:TBCS:2014,Pagkalos:TBCS:2014}, analog computing can be used to reduce energy footprint in computing and deep learning \cite{Owen:Science:2022,Gao:NatureComm:2022,Ambrogio:Nature:2023} and analog communications can be used in the nano- and/or bio-domain \cite{Daniel:2013ke,Chou:gc}. 

\section{Experimental evaluation}
\label{sec:evaluation}  

\subsection{Dataset}
We utilize a publicly available dataset \cite{Thuan2023} for diagnosing defects in ball bearing systems. The dataset covers 6 fault types: inner crack, outer crack, ball crack and pairwise combinations of these three types of faults. The dataset also contains data when faults are absent, i.e., in the healthy state. The data cover 5 bearing types (6204, 6205, 6206, 6207, and 6208) under 3 working conditions (0W, 200W, and 400W). 

\begin{figure*}[t]
    \centering
    \includegraphics[width=\textwidth]{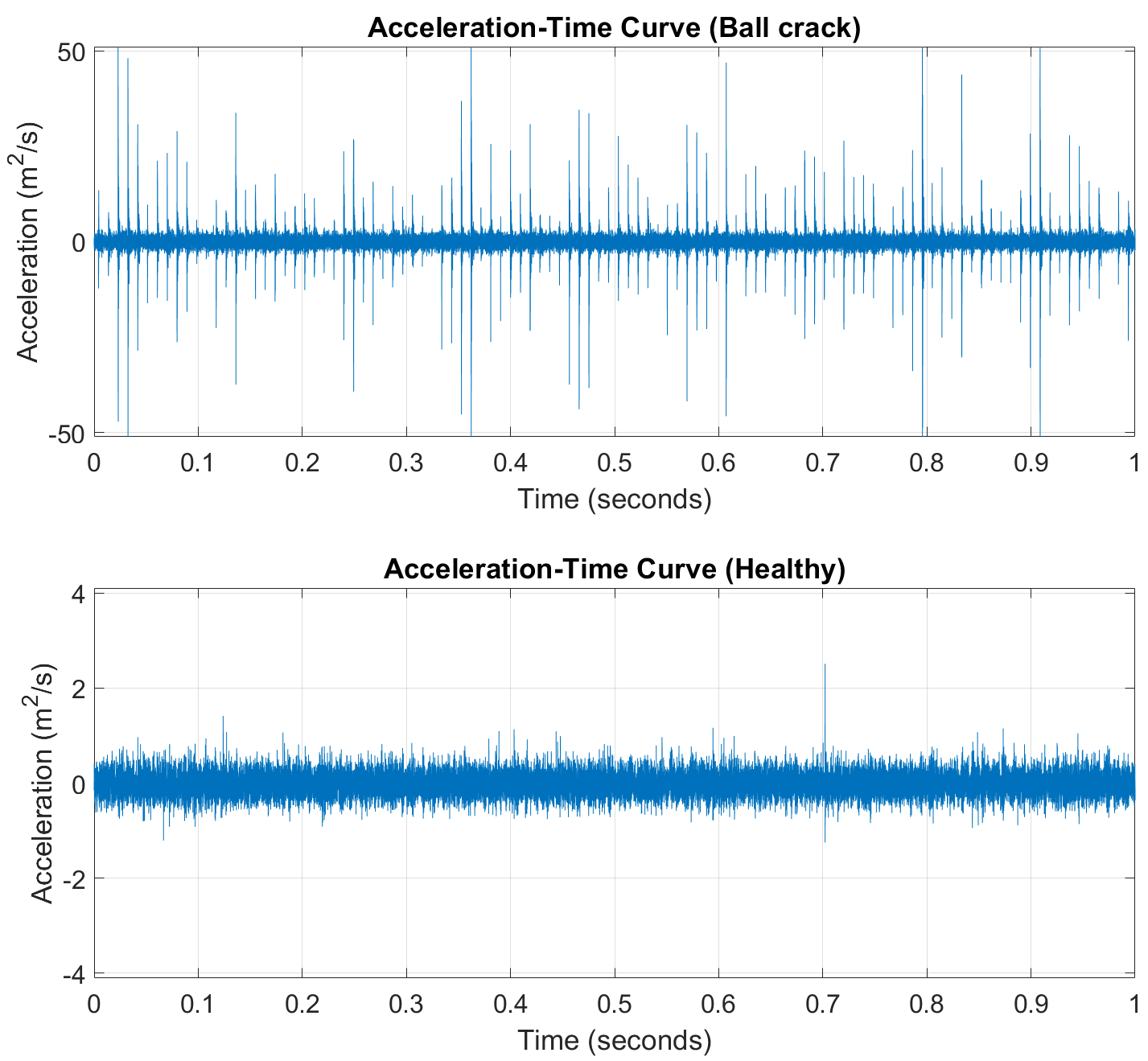}
    \caption{Sample acceleration traces for a faulty bearing with ball crack and a health bearing. }
    \label{fig:time:analysis}
\end{figure*}

\begin{figure}[h]
    \centering
    \includegraphics[width=\columnwidth]{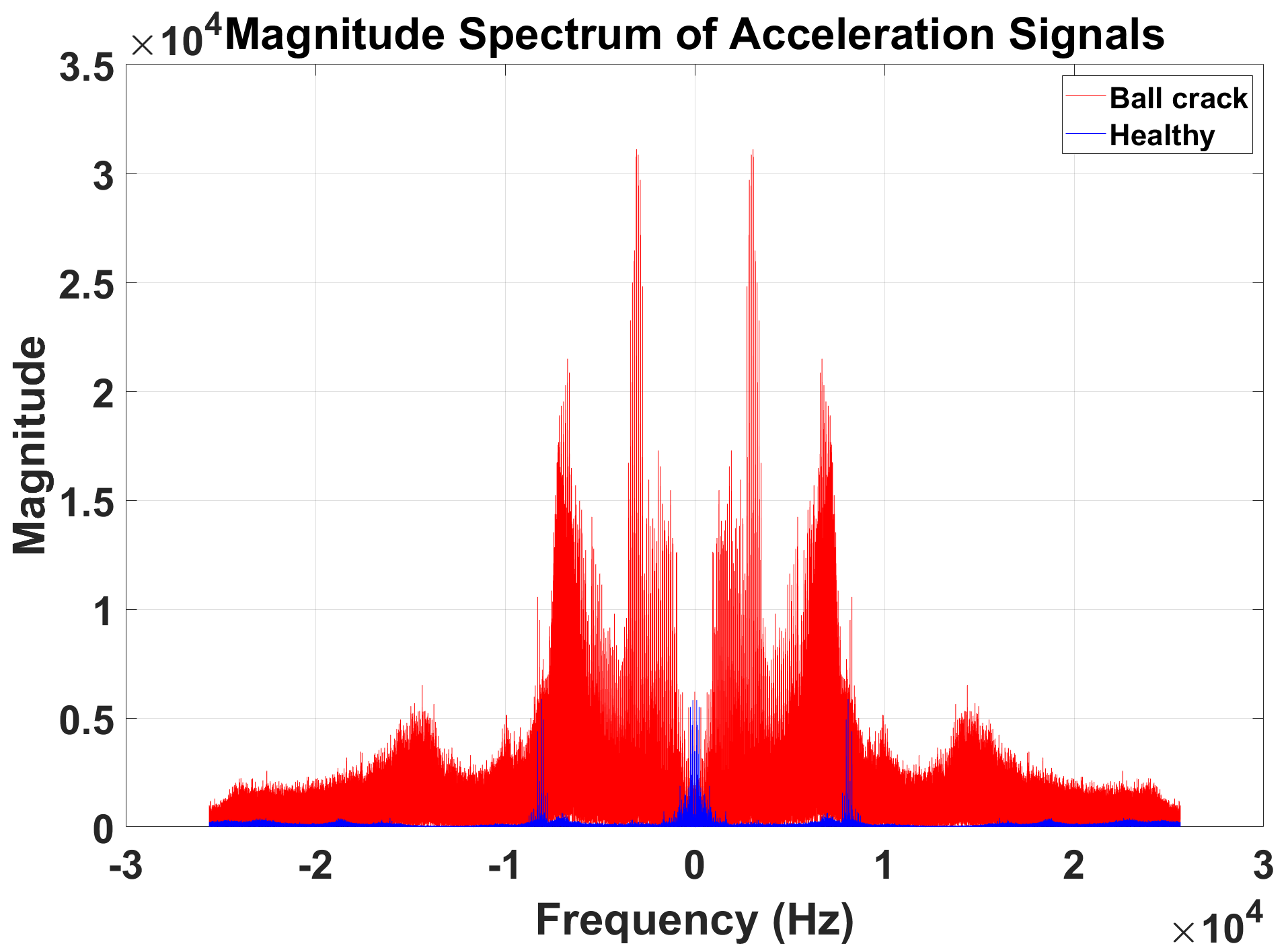}
    \caption{Frequency spectra of the acceleration signals for ball crack fault and healthy.}
    \label{fig:magni:acc}
\end{figure}

This dataset comprises of acceleration recordings collected over 10 seconds at a sampling rate of 51.2kHz. Each machine state, which is a type of fault or the healthy state, has 7 acceleration recordings. Fig.~\ref{fig:time:analysis} shows sample acceleration recordings for the healthy state and for ball crack. Note that we have plotted only one second of data so that the temporal features can be seen. It can be seen that these acceleration have different amplitudes and frequency contents. Fig.~\ref{fig:magni:acc} plots the magnitudes of the frequency spectra for the ball crack and healthy acceleration states. It can be seem that the frequency spectrum of healthy state has strong energy components in two frequency ranges while that of the ball crack state has energy distributed in a much wider frequency range. 


\subsection{Classification of faults}
In this paper we consider the detection problem to differentiation between ball crack fault and healthy state.

In this paper, we will use one PEH to differentiate between the faulty state and the healthy state. We use four different designs of PEHs by adjusting the thickness of the PZT layer. In this study, we use 4 thicknesses: 0.35mm, 0.4mm, 0.45mm and 0.5mm. The FRF of the four PEHs are shown in Fig.~\ref{fig:peh:frf}. 


We choose the integration time $T$ to be 3s. In order to obtain vibration signals that last for 3 seconds, we divide the first 9 seconds of each 10-second vibration signal recording into 3 non-overlapping segments of 3 seconds each. This gives altogether 21 3-second vibration signals. 

\begin{figure}[t]
    \centering
    \includegraphics[width=.9\columnwidth]{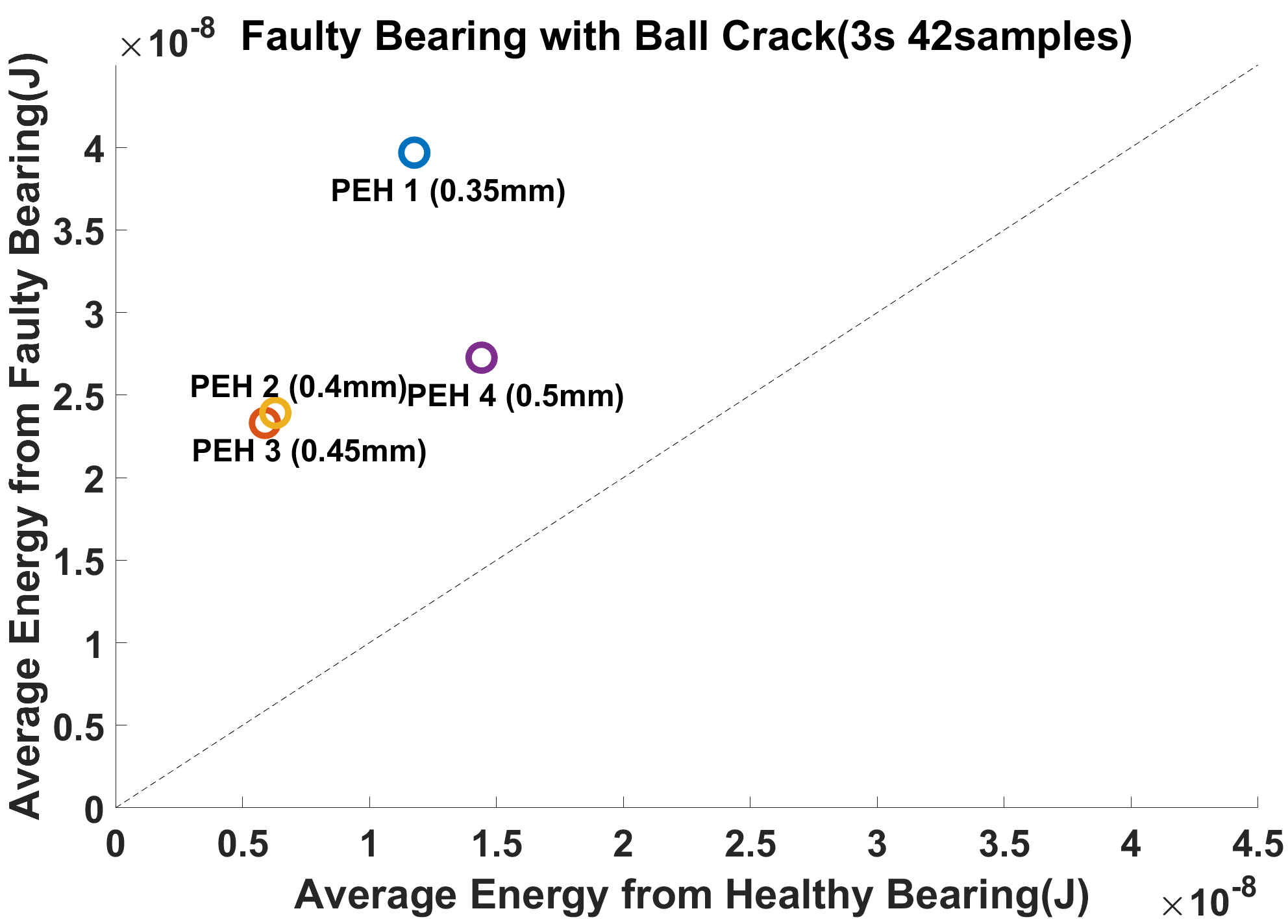}
    \caption{Ball crack versus healthy. $T$ = 3s.}
    \label{fig:energy:distribution:1F3s}
\end{figure}

We first examine whether it is possible to use the integrated harvested energy of one PEH to differentiate between the ball crack fault state and the healthy state. Fig.~\ref{fig:energy:distribution:1F3s} plots the average integrated harvested energy over a duration of $T = 3$ in the faulty state against that in the healthy state. Each marker in the plot is for a PEH of a specific thickness. 
The dashed lines in the figure are at 45$^{\circ}$ so if the markers are far away from the line, then it means there is a better chance to differentiate between the fault and the healthy states. It can be seem that the PEHs with thicknesses 0.35mm and 0.45mm are farther away from the 45$^{\circ}$ line. 


We investigate the classification accuracy by using a feature vector formed from one PEH. We use the $k$-Nearest Neighbors ($k$NN) algorithm for classification. We choose $k$ = 3. We use 80\% of the data as the training set and the remaining 20\% as the validation set. 

\begin{center}
    \begin{figure}[h]
        \centering
        \includegraphics[width=\columnwidth]{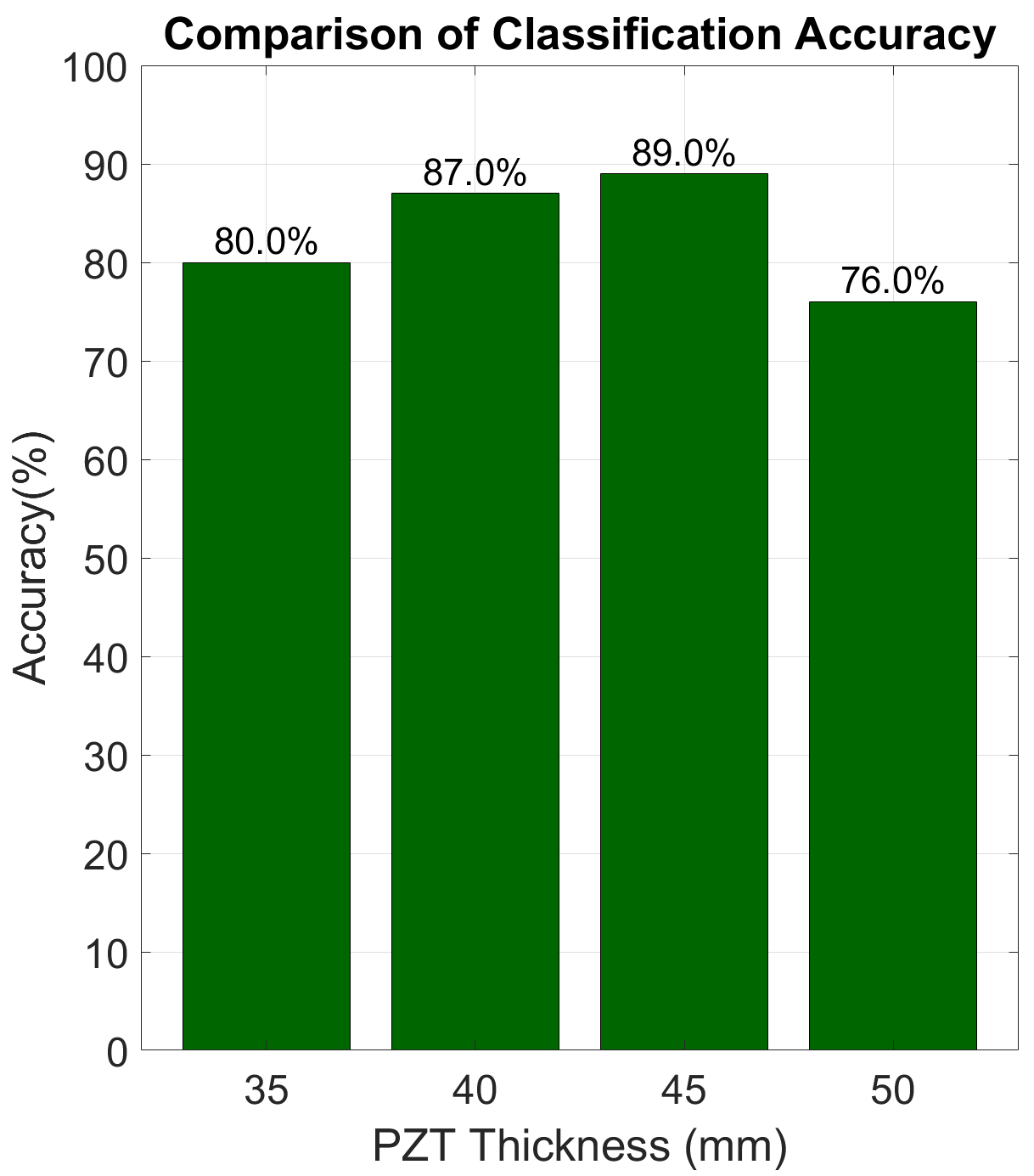}  
        \caption{Detection accuracy of the four PEHs.}
        \label{fig:comparison:accuracy}
    \end{figure}
\end{center}

Fig.~\ref{fig:comparison:accuracy} shows the detection accuracy for outer crack (left) and ball crack (right) for different choices of PEH thickness (hence frequency response) and integration time $T$. We are able to achieve an accuracy of 89\% for detecting the ball crack faults by using a PEH with 0.45mm thickness. 

A choice of using $T$ = 3 means that we use a sampling frequency of 0.33Hz to relay the integrated energy value to the server. In comparison with the original data which was sampled at 51.2kHz, this is a 4 orders-of-magnitude reduction. 

\begin{remark}
The architecture in Fig.~\ref{fig:current_vs_proposed} assumes that all the signal samples are transmitted to a server where the features will be estimated. One may ask whether there is any advantage on doing the Fast Fourier Transform (FFT) on the device side (where the computation will be done in a block between ADC and wireless transmission) and then only transmit the relevant Discrete Fourier Transform (DFT) coefficients to the server. This method will certainly reduce the amount of data transmitted wirelessly, however, one still has to pay for the energy cost of ADC as well as the energy cost for computing some DFT coefficient.  

\end{remark}

\section{Conclusions}
In this paper, we propose to use frequency selective piezo-electric harvesters to directly extract features for fault detection. The advantage is that we can drastically reduce the sampling requirement by 4 orders-of-magnitude. This reduction can reduce the energy consumption of ADC and wireless transmission, and therefore overcoming the energy consumption bottleneck in the existing data collection system for fault detection. Our future work is to investigate classification of all fault types, the design of PEHs etc. 

\bibliographystyle{IEEEtran}
\bibliography{main}

\end{document}